\documentclass[12pt,preprint]{aastex}

\shorttitle{Blueshifted [O III]}
\shortauthors{Boroson}

\received{2005 January 3}
\begin{document}


\title{Blueshifted [O III] Emission: Indications of a Dynamic NLR}


\author{Todd Boroson}
\affil{National Optical Astronomy Observatories, Tucson, AZ 85719}
\email{tyb@noao.edu}



\begin{abstract}
The [O III] $\lambda$5007 line is commonly used as an indicator of the systemic
redshift of AGNs.  Also, recent studies have used the width of this emission line 
as a proxy for the stellar velocity dispersion in the host galaxy.  This paper
calls both of these assumptions into question by analyzing a sample of approximately
400 AGN spectra from the first data release of the Sloan Digital Sky Survey.  These
spectra show that the low-ionization forbidden lines ([O II], [N II], [S II]) define a
consistent redshift, but that the peak of the [O III] line is blueshifted in 
approximately half of the AGNs with respect to that redshift.  For the sample 
studied here, the average
shift is 40 km s$^{-1}$, with the largest shift being over 400 km s$^{-1}$.  The
magnitude of this shift is found to be correlated with a number of properties, including
the width of the [O III] line and the Eddington ratio (L/L$_{Edd}$), derived from 
the luminosity and width of H$\beta$.  

\end{abstract}





\section{Introduction}

It is often assumed that the narrow emission lines in active galactic
nuclei (AGN) are indicative of some approximately stable, quiescent 
pool of gas that is being excited or illuminated by the central source.
Emission lines that originate in the narrow line region (NLR) are
commonly used to obtain the systemic redshift \citep{derobertis85, 
marzianietal96, richardsetal02} or as a
proxy for the stellar velocity dispersion \citep{shieldsetal03}.  Most 
often it is the [O III] $\lambda$5007 line that is used, as it is usually
the strongest and cleanest narrow line in the optical region of the spectrum.

The validity of the assumptions that the $\lambda$5007 line is a good indicator
of systemic redshift or stellar velocity dispersion rests primarily on extrapolation 
of studies of low luminosity objects, such as those by \citet{heckmanetal81}, 
\citet{vrtilek85}, \citet{nelsonandwhittle96}, and \citet{nelson00}.
However, even these studies of the NLR characteristics in lower luminosity
Seyfert galaxies suggest systematic departures from symmetry and stability
are not uncommon.
\citet{heckmanetal81} and subsequent studies found (1) that the $\lambda$5007 
line tends to be asymmetric, with a sharper falloff to the red than to the blue, 
and (2) that the redshift of the $\lambda$5007 line tends to be negative
compared with the systemic velocity derived from the stellar absorption lines.

Several recent studies have explored some of the characteristics of the [O III] $\lambda$5007
line in AGNs.  \citet{zamanovetal02} studied a sample of about 200 low redshift
AGN, most of which were luminous enough to be considered QSOs.  Under the assumption
that the redshift derived from the broad H$\beta$ line is the systemic value, they found 
that the $\lambda$5007 line gave a consistent redshift 90\% of the time, but that, among
the outliers, blueshifts were three times as common as redshifts.  They investigated
seven objects with the largest blueshifts, and found that they fell preferentially 
among the objects at the strong Fe II - narrow H$\beta$ end of the Eigenvector 1 
sequence \citep{bg92}.  In order to overcome the limitation of the assumption that
H$\beta$ gives the systemic redshift, they noted that for one of their blue outliers,
I Zwicky 1, other observations of the host galaxy give a redshift consistent with
the H$\beta$ line and not the [O III] line.  A consistent finding comes from the 
\citet{veronetal01} study of narrow line Seyfert 1 galaxies (NLS1s) in which it was 
found that approximately half of their sample had a broad, blueshifted [O III] 
$\lambda$5007 line in addition to the narrow component assumed to be at the systemic 
velocity.

\citet{boroson03} and \citet{nelson04} studied the [O III] $\lambda$5007 line in 
order to explore the utility of the width of this line as a surrogate for the stellar
velocity dispersion.  The \citet{boroson03} study looked at a sample of 107 QSOs
from the Sloan Digital Sky Survey (SDSS) Early Data Release (EDR), while the 
\citet{nelson04} study used new spectra of approximately 60 low-redshift PG QSOs.
The technique used in both studies was to plot the width of the $\lambda$5007 
line against black hole mass derived using the H$\beta$ broad 
line FWHM and the luminosity at $\lambda$5100 in the rest frame, L$_{5100}$.  
The tight correlation between black 
hole mass and stellar velocity dispersion \citep{tremaineetal02} in normal 
galaxies was the reference for the expected relationship.  Both studies found that 
the [O III] widths show some correlation with black hole mass, but the scatter 
is large.  \citet{nelson04} found that the more asymmetric [O III] lines have 
widths that are closer to those predicted by the \citet{tremaineetal02} fit.  That
study also found a correlation between [O III] line asymmetry and Eigenvector 1,
in that objects with greater asymmetry were more likely to be at the strong Fe II
end of the sequence, presumably correlating with higher L/L$_{Edd}$.

The study reported here is an effort to explore in greater detail and more quantitatively
the behavior of the [O III]
$\lambda$5007 line, by analyzing a larger sample of objects in which the systemic
velocity can be unambiguously determined.  By combining a number of measurements
of this line with other observed and derived properties, a better understanding
of the geometric and kinematic structure in which the [O III] emission arises may
emerge.  A cosmology with H$_0$ = 71 km s$^{-1}$ Mpc$^{-1}$, $\Omega_M$ = 0.27 and 
$\Omega_{\Lambda}$ = 0.73 is used throughout this paper.

\section{The Sample}


The sample of QSOs was drawn from Data Relase 1 (DR1) \citep{abazajianetal03} 
of the Sloan Digital
Sky Survey. The parent sample, which numbered about 3300, included all 
spectroscopically observed objects identified as QSOs having a redshift less 
than 0.5.  From this list, those with spectroscopic signal-to-noise less
than 15 in the R band (SN\_R in the FITS header) were removed, leaving 804 objects. 
The redshift distribution and absolute magnitude distribution (in the I band) for 
the sample are shown in figure 1.  The spectra were corrected to rest wavelengths,
adopting the SDSS value of the redshift stored in the header of each object.

These 804 spectra were inspected individually in the regions around the [O II] 
$\lambda$3727 doublet, the [N II] $\lambda$6584 line, and the [S II] $\lambda$6717 
and $\lambda$6731 lines.  When visible, the wavelength of each of these
lines was measured, by fitting a gaussian to the top half of the line.  These
wavelengths were then converted to redshifts, using the vacuum wavelengths
3728.30\AA, 6585.27\AA, 6718.29\AA, and 6732.67\AA \ as the rest values for the lines.
At least one of these lines could be measured in 654 of the 804 objects and at
least two of the lines from different ions could be measured in 399 of the 
objects. 

Figure 2 shows the distributions of relative redshifts for each of the three ions.
The histograms show the redshift of the identified line (or the average of the two 
[S II] lines) relative to the redshifts of any of the other lines that were measured. 
The average relative redshifts are 13.0, -11.2, and -3.6 km s$^{-1}$ for [O II],
[N II], and [S II] respectively.  The corresponding standard deviations of the
distributions are 39, 32, and 25 km s$^{-1}$.  It is likely that the [N II] measurements
are shifted a bit to the blue (negative velocities) because this narrow line typically
lies on the red wing of the broad H$\alpha$ line.  Part of the apparent shift of the 
[O II] distribution to positive velocity may be due to the fact that, in most cases, its
redshift is being calculated relative to the anomalously blue [N II] line.  An additional
factor which may also account for the greater width of the [O II] distribution is that
this `line' is actually a doublet with a separation between components of about 2.7A.
An average value for the centroid of the blended components is undoubtedly not a good guess
in some cases.

Despite the slight differences in mean and width of the distributions, they are all
roughly symmetric, and it is likely that these lines indicate the systemic velocity
of the galaxy.  All of these lines from singly ionized species have low critical
densities, and so they arise either in the outer parts of the 
narrow line region (NLR) or even further out, in the interstellar medium of the 
host galaxy itself.  The 3 arcsecond diameter of the SDSS fibers corresponds to a 
physical diameter of about 10.5 kpc at the sample median redshift of 0.22. The mean 
of all visible low ionization lines in each object 
was adopted as its systemic redshift, and the spectra were adjusted to reflect 
rest wavelengths using these new, small offsets.

From these corrected spectra, the wavelength of the [O III] $\lambda$5007 line was 
measured in each object in a manner similar to that in which the low ionization 
lines were measured.  These [O III] wavlengths were converted to a relative velocity 
using a vacuum rest wavelength of 5008.24\AA \ for this line.  Figure 3 shows the 
distribution of these [O III] redshifts relative to the systemic redshift calculated
from the low ionization lines.  The solid histogram represents all the objects for
which a low ionization systemic redshift could be determined and the [O III] 
$\lambda$5007 wavelength could be measured (638 objects; the total [O III] sample).  
The dotted 
histogram represents only
the objects for which low ionization redshifts could be determined from two or more
ions (399 objects; the high confidence [O III] sample).  The bottom panel of this figure 
shows the sum of the redshift
distributions for the three low ionization lines, in each case relative to the one or
two other lines measured.

The low ionization line distribution is well fit by a gaussian with a mean of zero
(by construction) and a width ($\sigma$) of 35 km s$^{-1}$, a fraction of which is due
to the slight shift between the distributions.  The [O III] distribution,
however, has a broad shoulder on the blueshifted side, with a sprinkling of objects
out to -400 km s$^{-1}$.  Approximately half the objects might be considered to be
in a distribution like that for the low ionization lines with the other half on the
blueshifted side.  For the total [O III] sample, 135 (21\%) of the 638 values have
positive (redshifted) relative velocities; this number is 93 (23\%) of 399 values for 
the high confidence sample.
Note that both the total [O III] sample and the high confidence [O III] sample are
biased within the original sample of DR1 objects toward the low redshift and low
luminosity end.  This bias is due to the fact that the low ionization narrow lines 
are more commonly found in the lower luminosity objects.  The dashed and dotted lines 
in the two panels of Figure 1 show the
distributions for the total [O III] and high confidence [O III] samples respectively.

\section{Properties of Objects with Different [O III] Redshifts}

In order to compare the properties of AGN with different [O III] velocity shifts, four 
subsamples were constructed.  The subsamples are composed of about a dozen objects
from the high confidence sample with [O III] velocities from the most blueshifted 
extreme of the distribution (the
VB sample), the moderately blueshifted region (the B sample), the region in which
the [O III] velocities are in good agreement with the systemic redshift (the N sample),
and the most redshifted extreme of the distribution (the R sample).  

For each object in these subsamples, the Fe II emission was measured and subtracted 
using the template and the technique developed in \citet{bg92}.  The continuum 
was then subtracted and the FWHM of the H$\beta$ line and the FWHM and asymmetry 
index A20 \citep{whittle85} of the [O III] 
$\lambda$5007 line were measured.  Table 1 lists the properties of each object in the
subsamples, including the absolute magnitude in the I band (from the I `PSF' magnitude 
listed in the SDSS DR1 database), the measured shift in the [O III] $\lambda$5007 line,
the equivalent width of the Fe II multiplets between $\lambda$4434 and $\lambda$4684,
the FWHM of H$\beta$ and [O III] $\lambda$5007, and the A20 value and equivalent width
of the [O III] 
$\lambda$5007 line.  Also listed are values of log M$_{BH}$ and log L/L$_{Edd}$
computed from the FWHM H$\beta$ and the luminosity at $\lambda$5100 in the rest frame, 
which was calculated
from the R band PSF magnitude.  The [O III] FWHM values are corrected for the instrumental
resolution, about 150 km s$^{-1}$.  At the bottom of the table, average values of each
quantity, for each of the four subsets, are given.

Inititially, an attempt was made to use the tabulated H$\beta$ width values from the SDSS
database.  Comparison with the final measured values shows that they are not a reliable
indicator of what is conventionally used as the FWHM of this line.  Only about 50\% of the
database values are within 20\% of the measured widths.  The discrepancies arise from
(1) the inclusion of the narrow component in the automatic measurement, (2) the 
departure of the profile from a Gaussian, and (3) effects due to contamination by
other lines - particularly Fe II.

Aside from the [O III] velocity values that characterize each sample, the statistical
difference among the samples are limited to a few properties.  Among the B, N, and R 
samples, there are no significant differences in the distribution of properties measured.
The VB sample is differentiated from the other three samples in having narrower H$\beta$
lines and broader [O III] $\lambda$5007 lines.  The narrower H$\beta$ lines result in
the VB sample having smaller black hole mass and larger L/L$_{Edd}$ on average than
any of the other samples.  It is worth noting that the VB sample does not differ from the 
other samples in the distribution of [O III] line asymmetry, A20. 
 
\section {Discussion}

It was previously known that among the objects with blueshifted [O III], Narrow Line 
Seyfert 1s (NLS1s), which
have H$\beta$ FWHM values less than 2000 km s$^{-1}$ and are thought to be high
L/L$_{Edd}$ objects, are 
overrepresented.  \citet {zamanovetal02} studied a small sample of objects with 
extreme [O III] blueshifts, greater than 250 km s$^{-1}$ relative to the peak of
the H$\beta$ line, drawn from a heterogeneous
sample of 216 objects.  These authors found that the seven extreme objects all had
H$\beta$ FWHM values below 4000 km s$^{-1}$, with three of the extreme 
objects falling in the NLS1 regime.  About half of the sample from which
these objects were drawn has H$\beta$ FWHM values below 4000 km s$^{-1}$.

Both the derived black hole mass and the Eddington ratio (L/L$_{Edd}$) 
depend on the H$\beta$ FWHM and the luminosity.  Limiting the comparison to
just the VB and N subsamples, the average luminosities of the two are not
significantly different but the H$\beta$ widths are quite different - 2175 km 
s$^{-1}$ for the VB sample vs 4456 km s$^{-1}$ for the N sample.  This leads to
black hole masses substantially smaller and Eddington ratios substantially 
larger on average for the VB objects.  A comparison of the distributions, however,
shows that the difference in Eddington ratio is far more significant than that in
black hole mass (0.8\% vs 4.5\% probability from a student t-test).  

Table 1 also shows, however, that not all narrow H$\beta$, high Eddington ratio
objects have large [O III] blueshifts and that not all the objects with large
[O III] blueshifts have narrow H$\beta$ and high Eddington ratios.  In order to
investigate whether other properties are related to the distinction between VB
and non-VB objects, another subsample, consisting of 12 objects that appeared to be
NLS1s with [O III] velocities near zero, was defined.  The same H$\beta$ and 
[O III] characteristics were measured for this sample, called the N-NLS1 sample, and
these values are also given in Table 1.  Other than [O III] velocity, this N-NLS1 
subsample has properties indistinguishable from the VB sample, except for the width
of the [O III] line, which has an average of 335 km s$^{-1}$, consistent with the
N value of 328 km s$^{-1}$, and inconsistent with the VB value of 523 km s$^{-1}$.
This is in contrast to the smaller average black hole masses calculated for the VB 
(and N-NLS1) objects, and therefore in conflict with the idea that the [O III] width 
is indicative of the stellar velocity dispersion, which is, in turn, correlated with 
the black hole mass. Thus, one could characterize the situation with two statements:
(1) objects with higher Eddington ratio are more likely to have large [O III] blueshifts; 
and (2) objects with large [O III] blueshifts have anomalously broad [O III] emission 
lines.

This result is shown graphically in Figure 4, in which black hole mass is plotted against the 
log of the [O III] line width ($\sigma$) for the separate subsamples.  Each subsample
is plotted as a different symbol, and the mean values for each subsample are plotted as
large symbols of the same type.  The line in Figure 4 is the \citet{tremaineetal02}
relation between black hole mass and stellar velocity dispersion. It is apparent that 
the VB subsample (solid circles) have similar black hole masses to the N-NLS1 subsample 
(open circles), but that the [O III] widths are much larger for the VB objects, which 
lie substantially below or to the right of the line.  It is interesting that the
N-NLS1 subsample also lies slightly below or to the right of the line, suggesting the
possibility of a similar, though smaller discrepancy between stellar velocity dispersion
and [O III] line width.

It has been claimed that objects with large [O III] blueshifts also are more likely 
to show blue asymmetries in that line.  That does not appear to be the case in this
sample.  There is no statistically significant difference in the distribution of
A20 values between any of the subsamples.  Only six of the 59 A20 values do not show 
a blue asymmetry, and, of these, two indicate no asymmetry in either direction.  
It appears that blue asymmetries are equally dominant in objects of all types,
both with and without [O III] blueshifts.  

The distribution of equivalent width of the [O III] $\lambda$5007 line among subsamples
is also worth noting.  While there is no formal difference between the distributions within,
admittedly small, subsamples, five of the 12 objects in the N sample have equivalent
widths larger than the largest object in the VB sample.  Thus, the N sample is
characterized by a much broader distribution (standard deviation = 31 \AA) of 
$\lambda$5007 equivalent widths than is the VB sample (standard deviation = 6 \AA).
The distributions differ formally at the 90\% significance level.

Many previous studies have speculated on the cause of [O III] emission with
an extended blue wing, a broad underlying blueshifted component, or a blueshifted
line center.  The explanations include both outflow models and inflow models.  
Outflow models are motivated by the idea that objects such as NLS1s have characteristics
that would support radiation pressure as a mechanism for accelerating such flows.
Imaging observations \citep{ruizetal05}, which show the [O III] emission coming from
a biconical region in many cases, support this idea.
These models must posit some obscuring screen, depending on the assumed scale of
the outflow, that blocks the observers view of the material on the far side.  This
could be the `dusty torus' or it could be a larger, central plane of interstellar 
material in the host galaxy. Inflow models have the advantage that they produce the 
line asymmetry by mixing the obscuring material with the NLR clouds, thus, no 
additional structure that happens to match the size of the outflowing, emitting 
material is needed.  It is the material on the near side that is invisible, since 
the backsides of these clouds are being illuminated.  These models are supported 
by the idea that the emitting clouds in NLS1s are the most dense - as deduced 
from studies of lines arising from ions such as Fe II, Ca II, and Na I 
\citep{ferland89}.  Thus, both of these types of models are consistent with the
increasing frequency of [O III] blueshifts with increasing L/L$_{Edd}$.

Although the new findings (as well as previous investigations) do not clearly distinguish
between the two models, a number of interesting conclusions may be drawn.  Radial motion
of narrow line emitting clouds is present in almost all objects.  This is the only way
to account for the common occurrence of line peak blueshifts and the ubiquitous presence
of the line asymmetry.  In the inflow model, we never see the near side of the clouds on
the near side of the nucleus; in the outflow model, we never see the clouds on the far
side of the nucleus.  This latter point is interesting because it implies that the scale
of narrow line emission is not large relative to some obscuring plane in the center of
the active galaxy, whether that is a structure related to the accretion flow or it is the
central plane of the galaxy.  This also suggests that the [O III] width is not generally
a suitable surrogate for the stellar velocity dispersion, but rather, represents the 
radial velocity gradient in the outflow or inflow.  

Why have previous studies found the [O III] line width to be correlated with stellar
velocity dispersion?  One possible explanation is that the outflow (or inflow) of the 
narrow-line emitting material is affected by the black hole mass, either through  
gravity or through radiation pressure.  If the [O III] line width is indicative of
the maximum radial velocity, then it would be correlated with the black hole mass.  At
the same time, the stellar velocity dispersion is correlated with the black hole mass,
and the two widths appear to be correlated with each other.

The most apparent flaw in this picture comes from the objects with extreme [O III] blueshifts.  
These are predominantly objects with small black hole masses and yet they have very 
broad [O III] lines.  Furthermore, they are, in all other ways, identical to another 
group of objects, the N-NLS1 subsample.  Clearly another parameter is required.  One
possibility for this parameter is orientation; objects with larger blueshifts and
broader lines are those
in which the outflow is directed closer to the line of sight.  

Putting these pieces together yields the following picture:  At a given L/L$_{Edd}$,
outflow (or inflow) velocity is larger in objects with larger black hole mass.  At a 
given black
hole mass, outflow velocity is larger in objects with larger L/L$_{Edd}$.  In addition,
as one goes to larger L/L$_{Edd}$, the cone angle of the outflow decreases, and in the
objects with large L/L$_{Edd}$, one sees a range from small outflow
velocities (in objects in which the axis of the outflow is in the plane of the sky) to 
large outflow velocities (in which the axis is along the line of sight).  Whether this 
picture extends all the way to truly edge-on objects in which one sees both sides 
of the outflow, and consequently symmetric lines with twice the flux, is not clear given
the small number of objects studied here.  This model implies that the other parameters
that distinguish objects - H$\beta$ width, Fe II strength, [O III] equivalent width - 
are not primarily orientation dependent.

An interesting comparison can be made with the velocity distribution of the peak of the
C IV line, studied in higher redshift SDSS quasars by \citet{richardsetal02}.  Working 
with composite spectra generated from large samples of objects divided by C IV blueshift
relative to the Mg II line, these authors found that more blueshifted C IV lines 
occurred as a result of the red wing of the line being obscured.  They suggested that 
this obscuration is due to a central screen of finite geometrical thickness, which 
is optically thin seen face-on but becomes thicker with increasing angle.  Note that this 
is, in some sense, opposite to the scheme proposed here, in which the objects with the 
largest blueshifts are those seen closest to face-on.  In any case, the geometry and 
kinematics of the C IV-emitting clouds and the [O III]-emitting clouds may be different,
and so the explanations may not be inconsistent.  

\section {Conclusions}

Characteristics of AGNs with blueshifted [O III] lines have been reported previously.
However, conclusions have been based on small samples, with heterogeneous data, or
requiring dubious assumptions.  In this study, a large, uniform sample of AGNs from
the SDSS DR1 have been analyzed.  Objects with discrepant [O III] redshifts have been
identified and subsamples constructed to explore the properties of objects populating
different regions in the [O III] relative velocity distribution.  Analysis of these
samples yields the following conclusions.

1) A significant fraction, as many as half, of all AGNs show [O III] emission that is 
blueshifted to the point that the line peak is shifted by tens to hundreds of km s$^{-1}$.

2) This shift is a true shift relative to the systemic velocity, as determined from 
the low ionization narrow forbidden lines.

3) The AGNs with the largest blueshifted [O III] lines tend to have high L/L$_{Edd}$, using 
black hole
masses determined from their luminosities and H$\beta$ widths.  However, this is not
a tight correlation; there are high L/L$_{Edd}$ objects that do not show blueshifts, and
lower L/L$_{Edd}$ objects that do.

4) The single anomalous property common to all the objects with highly blueshifted [O III]
emission is the large width of the [O III] lines.  These average more than 50\% broader
than those in similar objects that have [O III] close to the systemic velocity.  

5) Conversely, the blueshifted [O III] lines do not tend to be any more asymmetric or
any weaker in strength than the non-blueshifted ones.  All the subsamples show a similar 
tendency to have a sharp red edge to the line and a blue edge that is shallow, or has 
a wing or a shoulder.

While these characteristics do not identify a unique model, simple arguments suggest 
that the maximum radial velocity of the [O III] emitting clouds is controlled by both 
black hole mass and Eddington ratio.  A radial velocity gradient is primarily responsible
for the line width, which then allows an orientation parameter to play a role.  The 
apparent correlation of [O III] line width with stellar velocity dispersion in 
previous studies may be due to the fact that both of these widths are somewhat dependent 
on black hole mass.  

\acknowledgments

I thank G. Richards, P. Hall, J. Shields, G. Shields, and A. Laor for 
helpful discussions.  Funding for the creation and distribution of the SDSS 
Archive has been provided by the Alfred P. Sloan Foundation, the Participating 
Institutions, the National Aeronautics and Space Administration, the National 
Science Foundation, the Department of Energy, the Japanese Monbukagakusho, and 
the Max Planck Society. The SDSS Web site is http://www.sdss.org. The Participating 
Institutions in the SDSS are the University of Chicago, Fermilab, the Institute 
for Advanced Study, the Japan Participation Group, the Johns Hopkins University, 
the Max Planck Institute for Astronomy, the Max Planck Institute for Astrophysics, 
New Mexico State University, Princeton University, the United States Naval 
Observatory, and the University of Washington.

\clearpage



\begin{figure}
\epsscale{.80}
\plotone{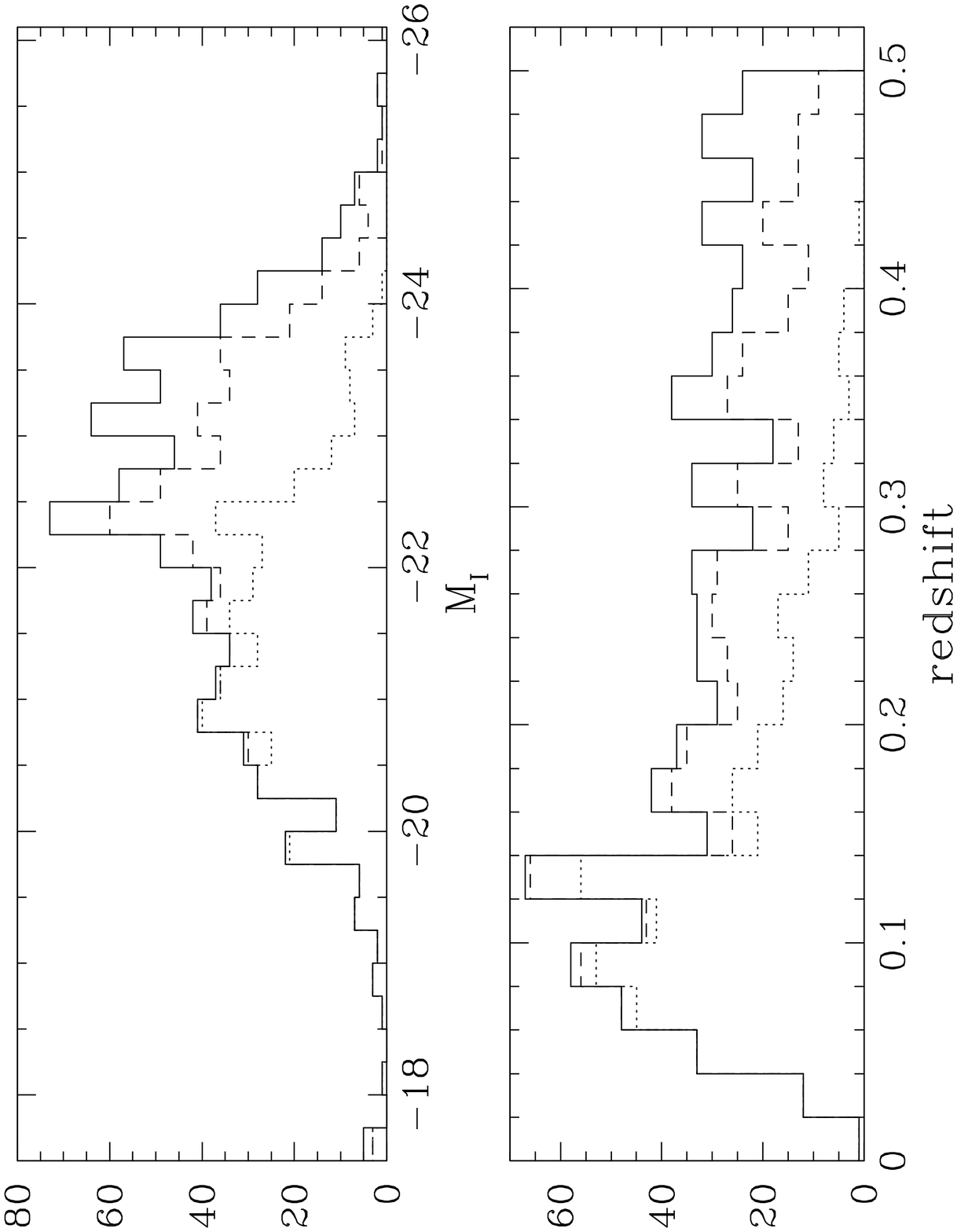}
\caption{Distribution of absolute magnitude in the I band (top) and redshift (bottom)
for the sample of 804 SDSS DR1 quasars studied.  The dashed line indicates those for
which any low ionization lines were measured.  The dotted line indicates those for
which lines from two or more low ionization species were measured.\label{fig1}}
\end{figure}

\clearpage


\begin{figure}
\plotone{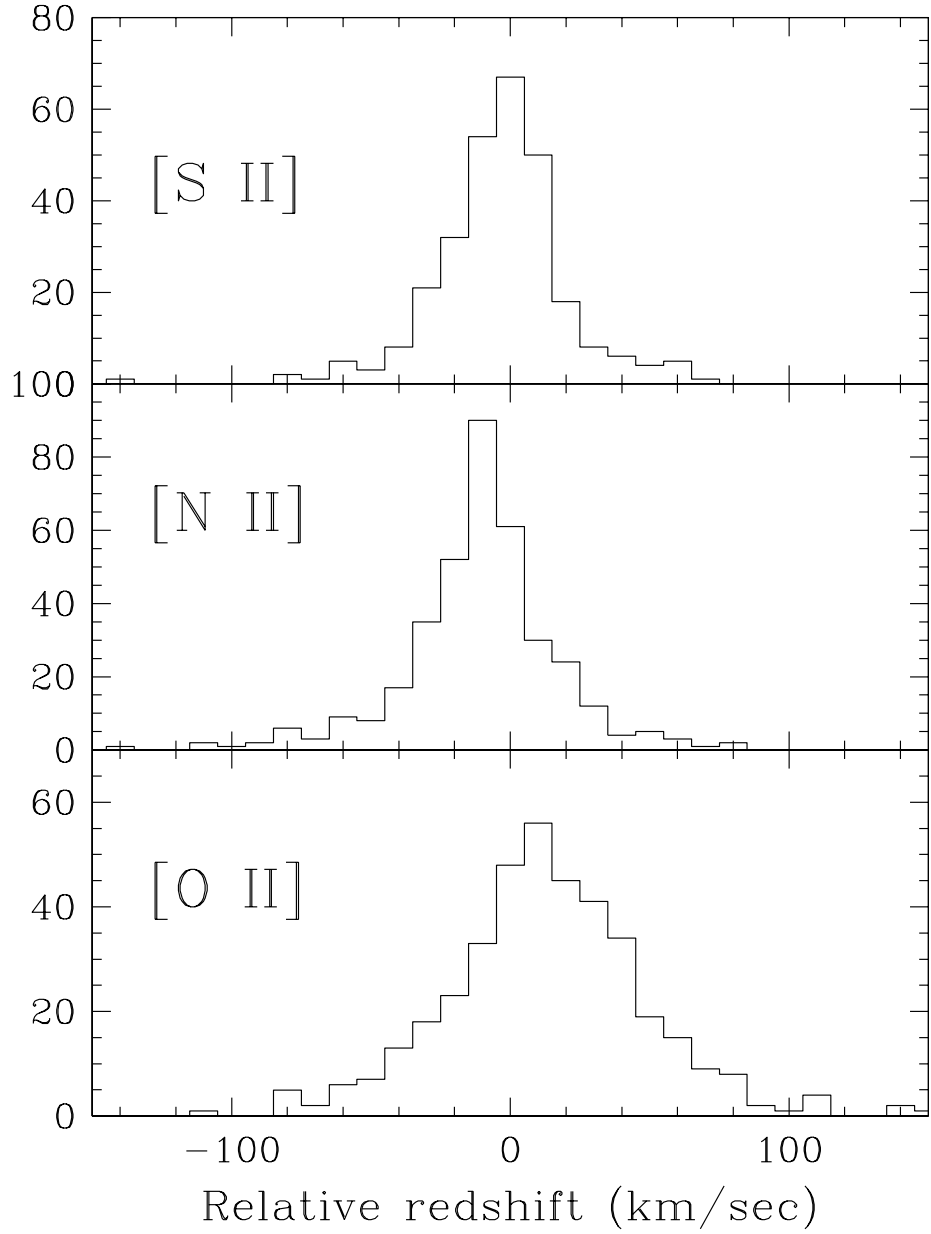}
\caption{For each of the three ions indicated, the distribution of redshift relative
to that of other low ionization lines.\label{fig2}}
\end{figure}

\begin{figure}
\plotone{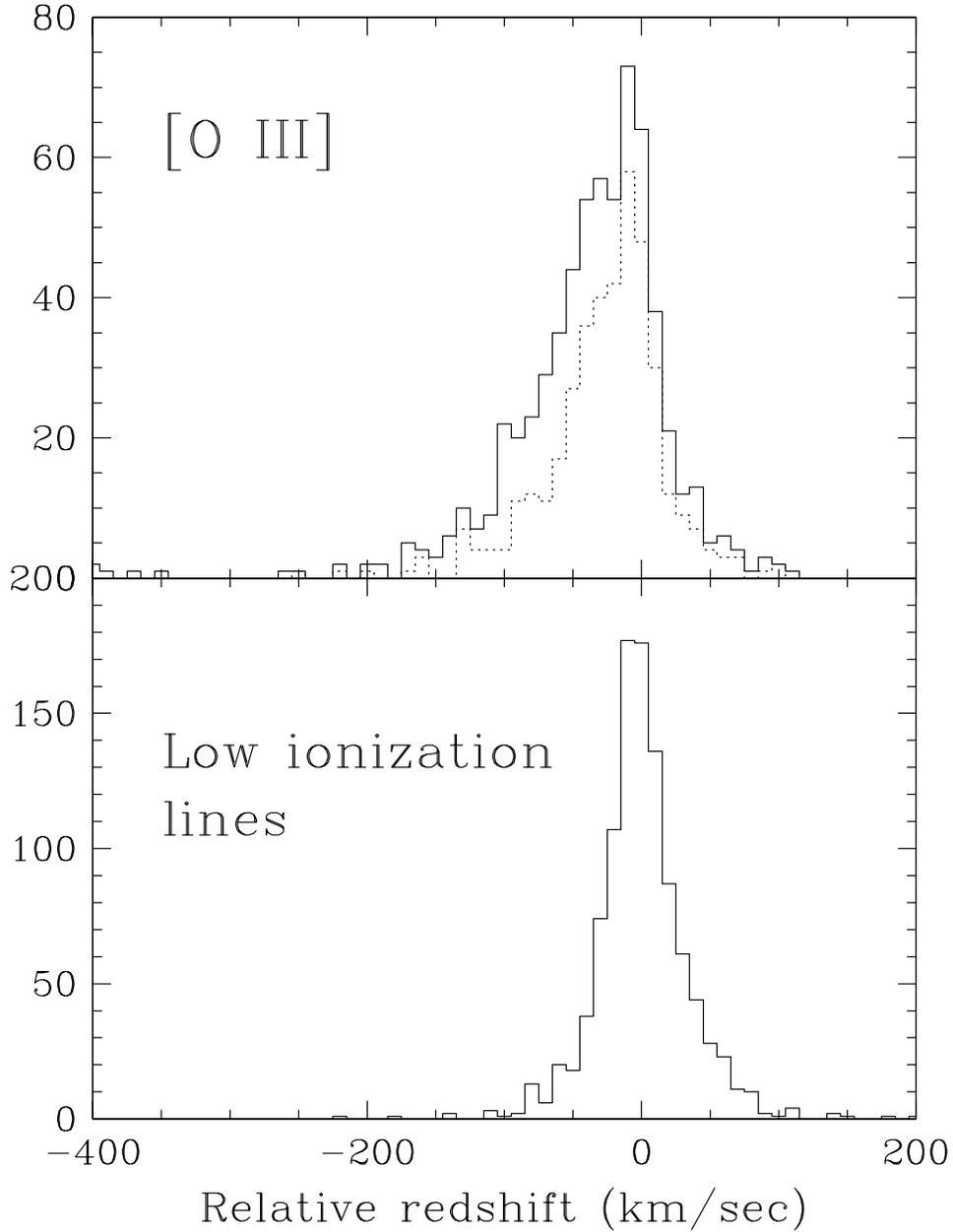}
\caption{Relative redshift distributions for all measurements of low ionization
lines (bottom) and for the [O III] $\lambda$5007 line (top). The dotted line in
the top panel shows the objects for which two or more low ionization species are
used to calculate the systemic velocity.}
\end{figure}

\begin{figure}
\plotone{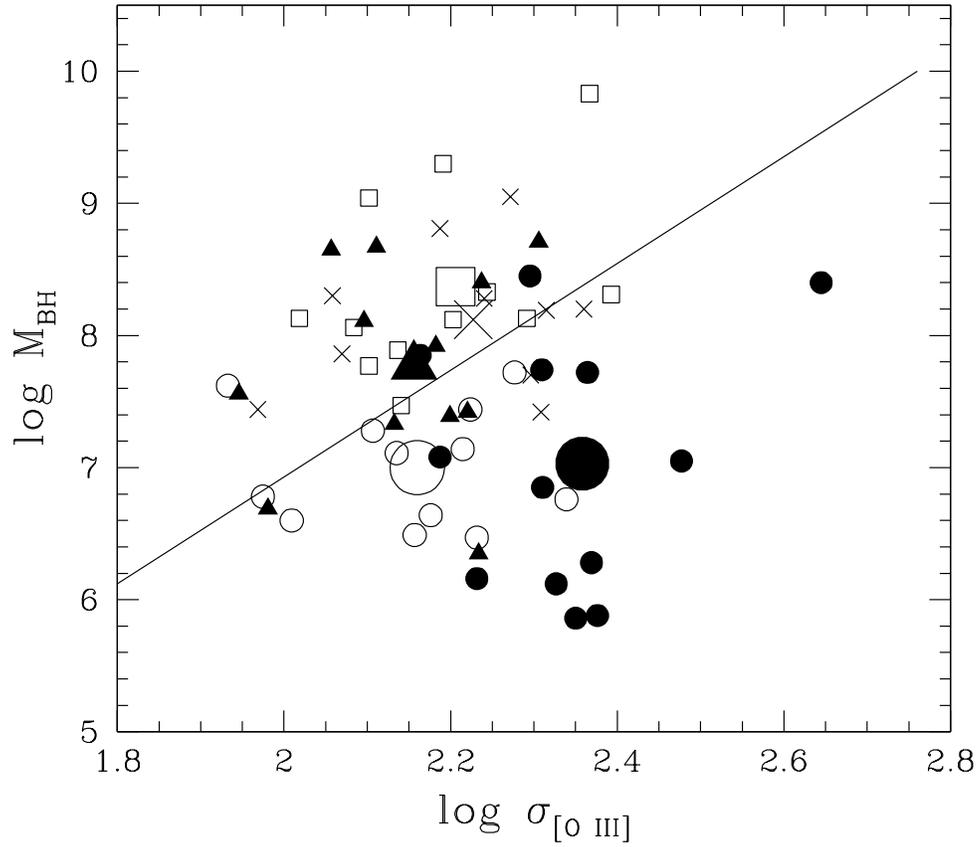}
\caption{Black hole mass plotted against the [O III] line width. The VB subsample is
plotted as solid circles, the B subsample is plotted as open squares, the N subsample
is plotted as solid triangles, the R subsample is plotted as crosses, and the N-NLS1
subsample is plotted as open circles.  The large symbols of each type show the position
of the mean values for that subsample. The line is the \citet{tremaineetal02} fit to 
the relation between black hole mass and stellar velocity dispersion.\label{fig4}}
\end{figure}






\clearpage

\begin{deluxetable}{clcccccccccc}
\tabletypesize{\tiny}
\rotate
\tablecaption{Properties of 5 AGN Subsamples\label{tbl-1}}
\tablewidth{0pt}
\tablehead{
\colhead{subsample} & \colhead{SDSS J} & \colhead{z} & \colhead{M$_I$} & \colhead{v$_{[O III]}$} & \colhead{EW Fe II} &
\colhead{FWHM H$\beta$} & \colhead{log M$_{BH}$} & \colhead{log L/L$_{Edd}$} &
\colhead{FWHM [O III]} & \colhead{A20 [O III]} & \colhead{EW [O III]}
}
\startdata
VB & 013521.68$-$004402.2 & 0.0985 & $-$20.91 & $-$252.8 & \phn 32 & \phn 1994 & 7.05 & \phs 0.32 & \phn 690 & \phs 0.248 & \phn 20.1 \\
VB & 011929.06$-$000839.7 & 0.0900 & $-$20.74 & $-$218.6 & \phn 38 & \phn \phn 715 & 6.12 & \phs 1.19 & \phn 488 & \phs 0.220 & \phn 15.5 \\
VB & 150521.92+014149.9 & 0.1580 & $-$21.31 & $-$165.9 & \phn 54 & \phn \phn 649 & 6.28 & \phs 1.38 & \phn 538 & \phs 0.052 & \phn 24.7 \\
VB & 154259.49+030653.3 & 0.0655 & $-$19.77 & $-$162.3 & \phn 17 & \phn \phn 712 & 5.86 & \phs 1.09 & \phn 515 & \phs 0.268 & \phn 15.5 \\
VB & 124635.25+022208.8 & 0.0482 & $-$20.44 & $-$159.3 & \phn 43 & \phn \phn 751 & 6.16 & \phs 1.15 & \phn 392 & \phs 0.208 & \phn 18.6 \\
VB & 032606.75+022539.9 & 0.1273 & $-$21.10 & $-$158.1 & \phn 58 & \phn \phn 508 & 5.88 & \phs 1.51 & \phn 547 & \phs 0.214 & \phn 12.0 \\
VB & 104153.60+031500.7 & 0.0934 & $-$20.45 & $-$133.6 & \phn 41 & \phn 2433 & 7.08 & \phs 0.08 & \phn 354 & \phs 0.041 & \phn 20.4 \\
VB & 131843.87+004732.2 & 0.3930 & $-$23.04 & $-$131.8 & 104 & \phn 4232 & 8.45 & $-$0.02 & \phn 454 & \phs 0.438 & \phn 16.6 \\
VB & 171829.01+573422.4 & 0.1008 & $-$21.16 & $-$130.6 & \phn 50 & \phn 1452 & 6.85 & \phs 0.63 & \phn 470 & \phs 0.229 & \phn 12.4 \\
VB & 082405.19+445246.1 & 0.2196 & $-$22.41 & $-$126.4 & \phn 32 & \phn 2439 & 7.72 & \phs 0.36 & \phn 532 & \phs 0.231 & \phn \phn 5.1 \\
VB & 132135.33$-$001305.8 & 0.0822 & $-$21.17 & $-$118.6 & \phn 21 & \phn 4498 & 7.85 & $-$0.35 & \phn 335 & \phs 0.051 & \phn 27.5 \\
VB & 171049.90+652102.1 & 0.3857 & $-$23.52 & $-$116.8 & \phn 73 & \phn 5351 & 8.40 & $-$0.33 & 1015 & \phs 0.228 & \phn 13.2 \\
VB & 091459.58+023510.9 & 0.2329 & $-$22.35 & $-$116.2 & \phn 21 & \phn 2543 & 7.74 & \phs 0.31 & \phn 469 & \phs 0.242 & \phn 11.2 \\
B & 113626.06+021144.2 & 0.1374 & $-$21.22 & \phn $-$55.1 & \phn 35 & \phn 2870 & 7.47 & \phs 0.05 & \phn 318 & \phs 0.133 & \phn 10.9 \\
B & 021218.32$-$073719.7 & 0.1738 & $-$22.45 & \phn $-$54.1 & \phn 74 & \phn 2446 & 8.12 & \phs 0.52 & \phn 367 & \phs 0.126 & \phn 15.1 \\
B & 022417.17$-$092549.3 & 0.3115 & $-$22.82 & \phn $-$53.9 & \phn 41 & \phn 8479 & 9.04 & $-$0.63 & \phn 291 & \phs 0.195 & \phn 31.7 \\
B & 162650.24$-$001731.8 & 0.2685 & $-$21.68 & \phn $-$53.3 & \phn 54 & \phn 5871 & 8.31 & $-$0.48 & \phn 568 & \phs 0.148 & \phn 19.4 \\
B & 012946.72+150457.3 & 0.3652 & $-$23.57 & \phn $-$53.3 & \phn \phn 0 & \phn 9346 & 9.30 & $-$0.63 & \phn 357 & \phs 0.000 & \phn 32.4 \\
B & 080644.41+484149.3 & 0.3701 & $-$23.68 & \phn $-$52.1 & \phn \phn 0 & 16868 & 9.83 & $-$1.14 & \phn 535 & \phs 0.495 & \phn 29.5 \\
B & 101911.34+015354.7 & 0.1886 & $-$22.21 & \phn $-$52.1 & \phn 36 & \phn 5699 & 8.33 & $-$0.44 & \phn 403 & \phs 0.000 & \phn 49.5 \\
B & 074948.26+345444.0 & 0.1318 & $-$22.28 & \phn $-$52.1 & \phn 71 & \phn 2920 & 7.77 & \phs 0.15 & \phn 291 & \phs 0.294 & \phn \phn 9.0 \\
B & 085635.86+521729.6 & 0.1443 & $-$21.15 & \phn $-$51.5 & \phn 25 & \phn 6279 & 8.13 & $-$0.64 & \phn 240 & \phs 0.442 & \phn 12.4 \\
B & 154751.94+025550.9 & 0.0980 & $-$21.47 & \phn $-$51.5 & \phn 35 & \phn 5868 & 8.13 & $-$0.56 & \phn 450 & \phs 0.285 & \phn 27.2 \\
B & 172033.62+580829.6 & 0.1597 & $-$21.91 & \phn $-$50.9 & \phn 14 & \phn 4353 & 8.06 & $-$0.22 & \phn 279 & \phs 0.136 & \phn 24.5 \\
B & 142245.79+630739.2 & 0.1608 & $-$22.20 & \phn $-$50.9 & \phn 71 & \phn 3194 & 7.89 & \phs 0.09 & \phn 315 & \phs 0.125 & \phn 10.6 \\
N & 083202.16+461425.7 & 0.0459 & $-$19.88 & \phn \phn $-$1.8 & \phn 23 & \phn 1773 & 6.69 & \phs 0.31 & \phn 220 & $-$0.040 & \phn \phn 8.5 \\
N & 130002.92$-$010601.8 & 0.3074 & $-$22.72 & \phn \phn $-$1.2 & \phn \phn 0 & \phn 5575 & 8.65 & $-$0.27 & \phn 262 & \phs 0.105 & \phn 63.7 \\
N & 013527.85$-$004448.0 & 0.0804 & $-$20.62 & \phn \phn $-$1.2 & \phn 21 & \phn 3524 & 7.42 & $-$0.23 & \phn 382 & \phs 0.126 & \phn \phn 2.9 \\
N & 035301.02$-$062326.4 & 0.0760 & $-$20.62 & \phn \phn $-$1.2 & \phn 14 & \phn 7190 & 8.11 & $-$0.82 & \phn 287 & \phs 0.159 & \phn \phn 7.7 \\
N & 134952.84+020445.1 & 0.0328 & $-$19.99 & \phn \phn $-$1.2 & \phn \phn 0 & \phn 3525 & 7.33 & $-$0.27 & \phn 312 & \phs 0.121 & \phn 91.4 \\
N & 150756.89+032037.3 & 0.1369 & $-$21.85 & \phn \phn $-$1.2 & \phn 14 & \phn 2204 & 7.39 & \phs 0.34 & \phn 364 & \phs 0.245 & \phn 44.0 \\
N & 090821.74+522103.0 & 0.0759 & $-$20.27 & \phn \phn $-$0.6 & \phn 22 & \phn 1089 & 6.35 & \phs 0.77 & \phn 394 & \phs 0.233 & \phn \phn 8.3 \\
N & 113633.09+020747.5 & 0.2390 & $-$22.65 & \phn \phn \phs 0.6 & \phn 31 & \phn 7456 & 8.71 & $-$0.61 & \phn 465 & \phs 0.329 & \phn \phn 9.6 \\
N & 134617.54+622045.5 & 0.1165 & $-$22.28 & \phn \phn \phs 0.6 & \phn 40 & \phn 8437 & 8.67 & $-$0.78 & \phn 297 & \phs 0.083 & \phn 47.7 \\
N & 170845.58+595716.2 & 0.2736 & $-$21.62 & \phn \phn \phs 0.7 & \phn \phn 0 & \phn 6254 & 8.40 & $-$0.52 & \phn 397 & $-$0.077 & \phn 72.6 \\
N & 123440.96+675213.8 & 0.2740 & $-$22.41 & \phn \phn \phs 1.8 & \phn 68 & \phn 1833 & 7.56 & \phs 0.64 & \phn 203 & $-$0.347 & \phn \phn 3.1 \\
N & 154606.97+034757.0 & 0.1274 & $-$21.61 & \phn \phn \phs 1.8 & \phn 89 & \phn 4612 & 7.92 & $-$0.35 & \phn 350 & \phs 0.463 & \phn 17.9 \\
R & 143847.54$-$000805.4 & 0.1040 & $-$20.83 & \phs \phn 33.5 & \phn 51 & \phn 5493 & 7.86 & $-$0.59 & \phn 270 & \phs 0.229 & \phn 10.5 \\
R & 003659.78+010544.4 & 0.1211 & $-$21.01 & \phs \phn 33.5 & \phn 53 & \phn 3053 & 7.44 & $-$0.04 & \phn 214 & \phs 0.132 & \phn \phn 7.9 \\
R & 110057.71$-$005304.5 & 0.3777 & $-$23.78 & \phs \phn 38.9 & \phn 70 & \phn 6399 & 9.05 & $-$0.27 & \phn 430 & \phs 0.315 & \phn 13.6 \\
R & 030639.57+000343.2 & 0.1071 & $-$21.58 & \phs \phn 46.1 & \phn 12 & \phn 2585 & 7.42 & \phs 0.15 & \phn 468 & \phs 0.175 & \phn 37.5 \\
R & 080322.48+433307.1 & 0.2761 & $-$22.32 & \phs \phn 52.7 & \phn 14 & \phn 4740 & 8.30 & $-$0.22 & \phn 263 & \phs 0.122 & \phn \phn 8.4 \\
R & 141556.85+052029.6 & 0.1263 & $-$22.01 & \phs \phn 52.7 & \phn 26 & \phn 5313 & 8.20 & $-$0.41 & \phn 527 & \phs 0.182 & \phn 59.4 \\
R & 030144.20+011530.9 & 0.0747 & $-$20.89 & \phs \phn 65.3 & \phn 28 & \phn 4187 & 7.70 & $-$0.32 & \phn 455 & \phs 0.323 & \phn 22.5 \\
R & 163631.29+420242.5 & 0.0610 & $-$20.54 & \phs \phn 68.9 & \phn 26 & \phn 8131 & 8.19 & $-$0.94 & \phn 475 & \phs 0.238 & \phn \phn 6.4 \\
R & 003847.97+003457.4 & 0.0805 & $-$20.93 & \phs \phn 71.9 & \phn 41 & \phn 8510 & 8.28 & $-$0.95 & \phn 400 & \phs 0.119 & \phn 21.9 \\
R & 015629.06+000724.4 & 0.3604 & $-$22.49 & \phs 104.8 & \phn 28 & \phn 7350 & 8.81 & $-$0.55 & \phn 354 & \phs 0.257 & \phn 16.5 \\
N-NLS1 & 081231.44+441620.8 & 0.2969 & $-$22.68 & \phn $-$10.8 & \phn 15 & \phn 2058 & 7.72 & \phs 0.56 & \phn 435 & \phs 0.269 & \phn 25.7 \\
N-NLS1 & 035107.60$-$052637.1 & 0.0691 & $-$20.72 & \phn $-$11.4 & \phn 59 & \phn 1356 & 6.76 & \phs 0.67 & \phn 502 & \phs 0.213 & \phn 12.2 \\
N-NLS1 & 123440.96+675213.8 & 0.2740 & $-$22.41 & \phn \phn \phs 1.8 & \phn 68 & \phn 1833 & 7.62 & \phs 0.67 & \phn 197 & $-$0.315 & \phn \phn 3.0 \\
N-NLS1 & 075333.82+385722.2 & 0.1468 & $-$21.29 & \phn \phn $-$2.4 & \phn 14 & \phn 1807 & 7.14 & \phs 0.48 & \phn 377 & \phs 0.013 & \phn 21.0 \\
N-NLS1 & 083202.16+461425.7 & 0.0459 & $-$19.88 & \phn \phn $-$1.8 & \phn 13 & \phn 1675 & 6.78 & \phs 0.42 & \phn 217 & \phs 0.000 & \phn \phn 8.0 \\
N-NLS1 & 151956.57+001614.6 & 0.1144 & $-$21.32 & \phn \phn \phs 4.2 & \phn 45 & \phn 1782 & 7.11 & \phs 0.48 & \phn 314 & \phs 0.158 & \phn 15.1 \\
N-NLS1 & 125227.32+032353.6 & 0.1328 & $-$21.65 & \phn \phs 15.6 & \phn 26 & \phn 1822 & 7.28 & \phs 0.53 & \phn 294 & \phs 0.473 & \phn 34.4 \\
N-NLS1 & 130713.25$-$003601.6 & 0.1700 & $-$22.49 & \phn \phn \phs 5.0 & \phn 35 & \phn 1931 & 7.44 & \phs 0.53 & \phn 385 & \phs 0.183 & \phn 69.8 \\
N-NLS1 & 090821.74+522103.0 & 0.0759 & $-$20.27 & \phn \phn $-$0.6 & \phn 22 & \phn 1089 & 6.47 & \phs 0.82 & \phn 392 & \phs 0.244 & \phn \phn 7.7 \\
N-NLS1 & 111407.35$-$000031.0 & 0.0727 & $-$19.90 & \phn \phn \phs 7.2 & \phn 11 & \phn 1230 & 6.60 & \phs 0.72 & \phn 235 & \phs 0.020 & \phn 15.8 \\
N-NLS1 & 135444.34+024039.2 & 0.1385 & $-$20.86 & \phn \phn \phs 4.8 & \phn 46 & \phn \phn 936 & 6.49 & \phs 1.02 & \phn 330 & \phs 0.327 & \phn 59.8 \\
N-NLS1 & 131008.01+634103.6 & 0.1926 & $-$22.35 & \phn \phs 10.2 & \phn 55 & \phn \phn 758 & 6.64 & \phs 1.34 & \phn 345 & \phs 0.200 & \phn \phn 7.0 \\

\cutinhead {Subsample Averages}
VB & Sample Average & 0.1612 & $-$21.41 & $-$153.2 & \phn 45 & \phn 2175 & 7.03 & \phs 0.56 & \phn 523 & \phs 0.205 & \phn 16.4 \\
B & Sample Average & 0.2091 & $-$22.22 & \phn $-$52.6 & \phn 38 & \phn 6183 & 8.37 & $-$0.33 & \phn 368 & \phs 0.198 & \phn 22.7 \\
N & Sample Average & 0.1488 & $-$21.38 & \phn \phn $-$0.2 & \phn 27 & \phn 4456 & 7.77 & $-$0.15 & \phn 328 & \phs 0.117 & \phn 31.5 \\
R & Sample Average & 0.1689 & $-$21.64 & \phs \phn 56.8 & \phn 35 & \phn 5576 & 8.12 & $-$0.41 & \phn 386 & \phs 0.209 & \phn 20.5 \\
N-NLS1 & Sample Average & 0.1441 & $-$21.32 & \phn \phn \phs 1.8 & \phn 34 & \phn 1523 & 7.00 & \phs 0.69 & \phn 335 & \phs 0.149 & \phn 23.3 \\
\enddata



\end{deluxetable}


\clearpage


\end{document}